\documentstyle[emulateapj]{article}
\def\lsim{\lower.5ex\hbox{$\; \buildrel < \over \sim \;$}}
\def\gsim{\lower.5ex\hbox{$\; \buildrel > \over \sim \;$}}
\lefthead{}
\righthead{  }

\def\lsim{\lower.5ex\hbox{$\; \buildrel < \over \sim \;$}}
\def\gsim{\lower.5ex\hbox{$\; \buildrel > \over \sim \;$}}

\begin{document}

\title{Higher order nonlinearity in accretion disks: QPOs of black hole and neutron star sources and
their spin}

\author{Banibrata Mukhopadhyay\altaffilmark{1}}
\altaffiltext{1}{Astronomy and Astrophysics Program, Department of Physics,
Indian Institute of Science, Bangalore 560012, India; bm@physics.iisc.ernet.in} 

\begin{abstract}

We propose a unified model to explain Quasi-Periodic-Oscillation (QPO), 
particularly of high frequency, observed from 
black hole and neutron star systems globally.
We consider accreting systems to be damped harmonic oscillators exhibiting
epicyclic oscillations with higher order nonlinear resonance to
explain QPO. The resonance is expected to be driven by the disturbance from 
the compact object at its spin frequency. The model explains various properties
parallelly for both types of the compact object. 
It describes QPOs successfully for ten different compact sources. 
Based on it, we predict the spin frequency
of the neutron star Sco~X-1 and specific angular momentum of 
black holes GRO~J1655-40, XTE~J1550-564, H1743-322, GRS~1915+105.

\end{abstract}


\section{Introduction}

The origin of Quasi-Periodic-Oscillation (QPO) and its properties are ill-understood phenomena in 
modern astrophysics. 
Observed QPO frequencies from black hole, neutron
star, as well as white dwarf systems range from milli Hertz (mHz) 
to kilo Hertz (kHz) with varying properties.
Among them most interesting and puzzling QPOs are the kHz ones. 
To date, more than $20$ accreting neutron stars in Low Mass X-Ray Binaries (LMXB) have
been found exhibiting such kHz frequency QPOs (van der Klis 2000; 2006).
Several black holes have also been found exhibiting the high frequency (HF)
QPOs of a few hundreds of Hz. The most puzzling cases are
those when a pair of QPO forms, particularly in neutron star systems, and QPOs in the pair
appear to be separated in frequency either 
by the order of the spin frequency of the neutron star, apparently for {\it slow rotators}, or
by half of the spin frequency, for {\it fast rotators}. 
An example of the later cases is 4U~1636-53 with the spin frequency $\nu_s\sim 580$ Hz, 
while of the former ones is 4U~1702-429 with $\nu_s\sim 330$ Hz.  
Moreover, in several occasions the frequency separation decreases
with the increase of one of the QPO frequencies. In black hole binaries also the HF QPOs 
have been found in pairs, but seem to appear at a $3:2$ ratio 
(Remillard et al. 2002) that favors the idea of a resonance mechanism behind their origin. 
The examples are GRO~J1655-40, XTE~J1550-564, GRS~1915+105
which exhibit the HF QPOs. 
The difference of the twin HF/kHz (van der Klis 2005)
QPOs between black holes and neutron stars might be due to presence of a rotating 
magnetosphere in the later cases imprinting directly the spin frequency on the 
oscillations of the disk. However, sometimes QPOs from neutron stars, e.g. Sco~X-1, also 
seem to appear at a $3:2$ ratio (Abramowicz et al. 2003).

Magnetically driven warping and precession instability are expected to occur in the 
inner region of an accretion disk (Lai 1999) 
by interaction between the surface current density in the disk and the stellar magnetic field. 
This produces motions at low frequencies, 
much lower than the orbital frequency accounting for the low frequency QPOs. 
This is similar to the Lense-Thirring precession shown by Stella \& Vietri (1998). The mHz
QPOs in some systems can be explained with magnetic warping process for strongly
magnetized neutron stars (Shirakawa \& Lai 2002). 
Titarchuk (2003), on the other hand, described QPO by the Rayleigh-Taylor instability 
associated with Rossby waves and rotational splitting. 
The effect of nonlinear coupling between
g-mode oscillations in a thin relativistic disk and warp was examined by Kato (2004)
for a static compact object.

A strong correlation between low and high frequency QPOs has been found by
recent observations (Psaltis, Belloni \& van der Klis 1999; Mauche 2002; Belloni, Psaltis
\& van der Klis 2002) of accretion disks around compact objects. 
Holding the relation over mHz to kHz range strongly
supports the idea that QPOs are universal physical processes, independent of the
nature of the compact object. Observing universal nature 
between the HF/kHz QPOs in black holes and neutron stars,
McClintock \& Remillard (2004) suggested that they are due to the orbital 
motion of the accreting matter near the marginally stable circular orbit.
Titarchuk \& Wood (2002) explained the correlation
in terms of a centrifugal barrier model 
(see also Titarchuk, Lapidus \& Muslimov 1998). 
Indeed, earlier by numerical simulation, 
Molteni, Sponholz \& Chakrabarti (1996) showed QPO in the black hole
accretion disks from the shock oscillation induced by the strong 
centrifugal barrier. 
Later, Chakrabarti \& Manickam (2000) verified this analyzing observational data. 

In early, it has been shown that QPOs may arise from nonlinear resonance phenomena 
in an accretion disk placed in strong gravity
%(e.g. Abramowicz \& Kluzniak 2001; 
(e.g. Kluzniak et al. 2004; Petri 2005; 
%T\"or\"ok et al. 2005; 
Blaes et al. 2007). 
This is particularly suggestive because of their variation in frequency, while we know 
that the varying frequency in a characteristic range is a well studied feature of 
nonlinear oscillators. 
Indeed accretion dynamics and the disk structure are determined by the nonlinear 
hydrodynamic/magnetohydrodynamic equations with the effective gravitational
potential which is not harmonic. 
Therefore, a nonlinear response is expected. 
The above authors argued for the resonance between epicyclic motions of accreting
matter around a rotating compact object resulting QPOs,
particularly the HF/kHz ones. 
However, the separation between vertical and radial epicyclic frequencies as a function of
radial coordinate increases mostly with the increase of either of the frequencies. 
Although the separation of
QPO frequencies in a pair observed from XTE~J1807-294 appears constant (Linares et al. 2005)
and from Aql~X-1 seems increasing (Barret, Boutelier \& Miller 2008) 
with the increase of either of the frequencies,
in most of the cases it is opposite (modeled by 
Stella \& Vietri 1999; Karas 1999) which contradicts the model based on 
the resonance between epicyclic motions. 
At radii smaller than the radius showing peak of 
radial epicyclic frequency, while the separation decreases with the increase of radial epicyclic frequency,
the frequency separation is too large, much larger than the spin frequency of the star.
Indeed, based on Cir~X-1 data, Belloni, M\'endez \& Homan (2007) argued against the model.
Therefore, one has to invoke new physics to describe multiple QPO features. 

In the present paper, we propose a {\it global} resonance model based on
higher order nonlinearity
to describe black hole and neutron star QPOs together. The model successfully describes 
important features including variation of the QPO frequency separation in a pair as a
function of QPO frequency itself.
On the mission of reproducing observed QPOs, the model predicts the
spin parameter (specific angular momentum or the Kerr parameter) 
of black holes as well as the spin frequency of a neutron star. 
The paper is organized as follows. 
In the next section,
we establish the model describing the underlying physics briefly. 
In \S 3 and \S 4 we reproduce
various properties of observed kHz/HF QPOs from six neutron stars and four black holes 
respectively, minimizing free parameters. Finally, we end with a summary in \S5.

\section{Model}

Wagoner (1999) and Kato (2004) investigated small oscillations of geometrically
thin accretion disks around a rotating compact object in the linear regime 
in the context of QPO. Later, Rezzolla et al. (2003) studied the radial
oscillations of twodimensional accretion tori in the Schwarzschild spacetime.
Kluzniak et al. (2004) investigated the response to a transient external
perturbation of an ideal gas torus around a nonrotating black hole or neutron star.
They found that the torus performs harmonic global oscillations in both the 
radial and the vertical directions with a variable amplitude in the later direction.
These effects indicate the existence of a nonlinear coupling into the system.

We know that a system of $N$ degrees of freedom has $N$ linear natural frequencies
denoting $\nu_1,\nu_2,\nu_3,......\nu_N$ (Nayfeh \& Mook 1979). Depending on the
order of nonlinearity to the system,
these frequencies are commensurable with certain 
relationships which may cause the corresponding modes to be strongly coupled and yield
an internal resonance. If the system is excited by a mode of frequency $\nu_*$, then 
frequencies might have the commensurable relationship exhibiting the resonance
\begin{eqnarray}
a\nu_*=\sum_{i=1}^N b_i \nu_i,
\label{frqrel}
\end{eqnarray}
(apart from all the primary and secondary resonance conditions $c\nu_*=d\nu_m$, 
where $\nu_m$ corresponds to the $m$th mode, $a,b_i,c,d$ are integers) such that
\begin{eqnarray}
a+\sum_{i=1}^N|b_i|=j,
\label{intrel}
\end{eqnarray}
where $j=k+1$ when $k$ is the order of nonlinearity.  
Certainly the order of nonlinearity determines the type of resonance conditions. 
For details, see Nayfeh \& Mook (1979).

Now we consider an accretion disk to be a system of higher degree nonlinearity 
and described as a damped oscillator with higher degree anharmonicity. 
%as chosen by Kluzniak et al. (2004) restricting with cubic anharmonicity. 
Then we consider the possible resonance therein. 
The resonance is expected to be driven by the combination of the strong disturbance by the
compact object with spin $\nu_s$ and other already existent (weaker) disturbances in
the disk particles (and hence disk) at the frequencies
of the radial ($\nu_r$) and vertical ($\nu_z$)
epicyclic oscillations given by (e.g. Shapiro \& Teukolsky 1983)
\begin{eqnarray}
%\nonumber
\nu_r=\frac{\nu_o}{r}\sqrt{\Delta-4(\sqrt{r}-a)^2},~
\nu_z=\frac{\nu_o}{r}\sqrt{r^2-4a\sqrt{r}+3a^2},
\label{epi}
\end{eqnarray}
where $\Delta=r^2-2r+a^2$, $r$ is radial coordinate expressed in unit of $GM/c^2$,
$a$ is specific angular momentum (spin parameter) 
of the compact object expressed in unit of $GM/c$, $G$ the Newton's gravitation 
constant, $c$ is speed of light, $M$ is mass of the compact object,
and $\nu_o$ is orbital frequency of the disk particles given by
%$2\pi\nu_o=\Omega=1/(r^{3/2}+a)$.
\begin{eqnarray}
2\pi\nu_o=\Omega=\frac{1}{r^{3/2}+a}\,\frac{c^3}{GM}.
\label{orb}
\end{eqnarray}

\begin{figure}
\epsscale{0.8}
%\hskip-5cm
\plotone{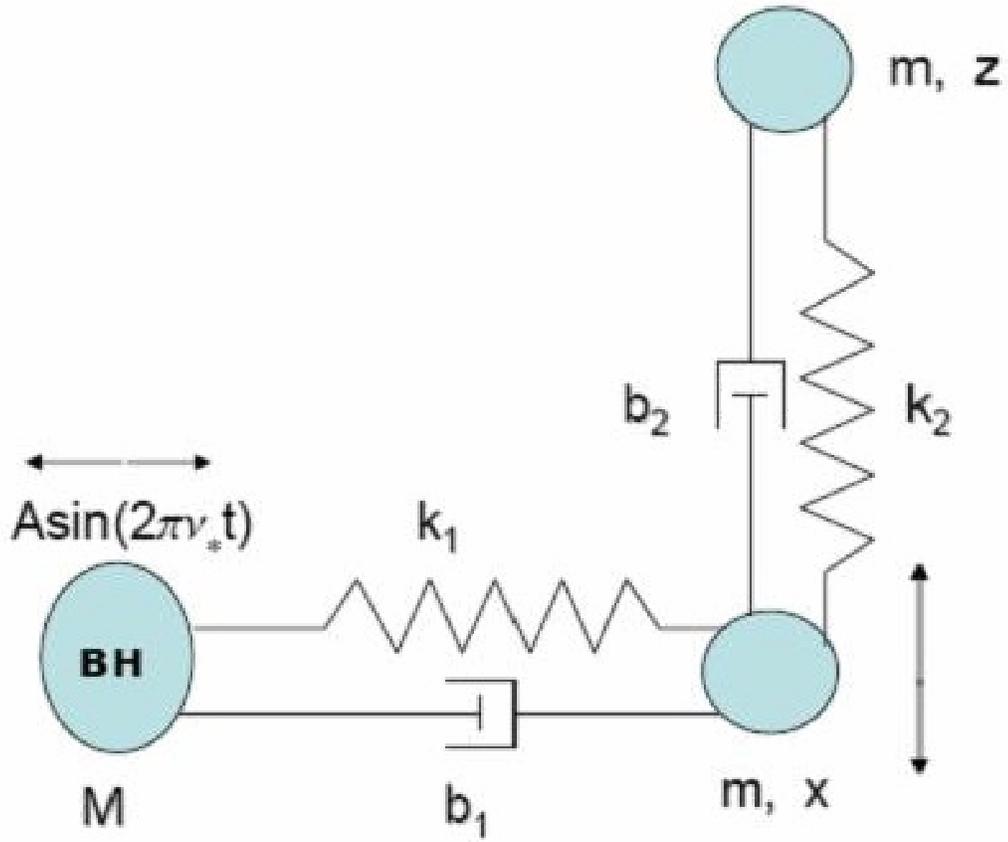}
\caption{ \label{figcart} Cartoon diagram describing coupling of various modes
in an accretion disk and the corresponding nonlinear oscillator. The oscillators 
describing by the spring constant $k_1$ and $k_2$ indicate
respectively the coupling of spin frequency $\nu_*$ of the compact object with mass $M$
to radial ($x$) epicyclic frequency
and vertical ($z$) epicyclic frequency of a disk blob with mass $m$, 
where $b_1$ and $b_2$ represent corresponding damping factors respectively. 
}
\end{figure}

We describe the system schematically in Fig. \ref{figcart}. It can be seen to have composed of
two oscillators, one in the radial direction (radial epicyclic motion) and other in the
vertical direction (vertical epicyclic motion), with different spring constants. Note that
spring constants here vary with radial coordinate.
The basic idea is that the mode corresponding to 
$\nu_s$ arised due to the disturbance created by rotation of the compact object
will couple to the ones corresponding to $\nu_r$ and $\nu_z$, already exist in the disk,
exciting new modes with frequencies $\nu_{r,z}\pm p\nu_s$ (Nayfeh \& Mook 1979),
where $p$ is a number, e.g.  $n/2$ when $n$ is an integer ($1,2$ for the present model). 
Now at a certain radius in the nonlinear regime where
$\nu_s/2$ (or $\nu_s$ in the linear regime) is close to $\nu_z-\nu_r$ (Nayfeh \& Mook 1979; 
Kluzniak et al. 2004),
if $\nu_s/2$ (or $\nu_s$) is also coincidentally close to the frequency difference
of any two newly excited modes, then a resonance may occur which locks the frequency difference
of two excited modes at $\nu_s/2$ (or $\nu_s$).
%The number $p$ is $k/2$ or $k$, when $k$ is
%an integer, depending on whether the effect is nonlinear or linear \cite{nm79,kluz04} respectively.
%While $k=1$ corresponds to the dominant interaction, modes with other $k$ are very weak to exihit
%any observable effect.

For a neutron star having a magnetosphere coupled with the disk matter, the mode with $\nu_s$ can 
easily disturb disk particles through the interaction between surface current density in the disk
and the stellar magnetic field and get coupled with the modes existing in
the disk. Ghosh \& Lamb (1979a,b) showed, close to the stellar surface, how does the 
magnetic field channel matter to the polar caps. Later, Anzer \& B\"orner (1980, 1983)
showed the influence of the Kelvin-Helmholtz instability. However, LMXB
have a very low surface field strength and a low critical fastness parameter
compared to highly magnetized neutron stars (Li \& Wickramasinghe 1997; Li \& Wang 1999).
Therefore, in this case, the strength of the disturbance is expected to be low and the accretion disk
usually extends close to the stellar surface (Popham \& Sunyaev 2001).

For a black hole, on the other hand, energy and angular momentum can be transferred
with the magnetic field, associated with its disk, threading the horizon 
and connecting to the surrounding environment
which may be looked as a variant of Blandford-Znajek process (Blandford \& Znajek 1977).
In fact, this kind of magnetic coupling process was already supported by XMM-Newton observations
(e.g. Wilms 2001; Li 2002). This confirms the possible excitation of new modes in the disk around
a spinning a black hole.

Note, as remarked by Karas (1999) in a similar context, 
that the disk particle size should be finite to visibly
modulate the observed radiation due to the resonance for a long period. However, shearing 
effect destroys large particles. Abramowicz et al. (1992) suggested that formation of vortices
in accretion disks could help in surviving finite size particles which was verified
by Adams \& Watkins (1995). While the origin of vortex is not fully 
understood yet, finite amplitude threedimensional secondary perturbations 
may form elliptical vortices in disks surviving for a long period 
(Mukhopadhyay, Afshordi, Narayan 2005; Mukhopadhyay 2006). In fact, Mukhopadhyay (2008),
based on it, argued for the origin of turbulent viscosity in accretion disks.

Therefore, we rewrite (\ref{frqrel}) for an accretion disk as
\begin{eqnarray}
(n-m+1)\nu_s=b_1\nu_r+b_2\nu_z
\label{frqrelac1}
\end{eqnarray}
which leads to 
\begin{eqnarray}
\nu_r+\frac{n}{2}\nu_s=\nu_z-\frac{\nu_s}{2}+\frac{m}{2}\nu_s
\label{frqrelac2}
\end{eqnarray}
with $-b_1=b_2=2$ and $m,n$ are integers. Now we propose the higher and lower QPO frequencies
of a pair respectively to be
\begin{eqnarray}
%\nonumber
\nu_h=\nu_r+\frac{n}{2}\nu_s,\,\ \nu_l=\nu_z-\frac{\nu_s}{2}.
\label{qpofrq}
\end{eqnarray}
%where $\nu_h$ and $\nu_l$ are the higher and lower QPO frequency respectively of a pair.
Hence, from (\ref{intrel}) and (\ref{frqrelac2}) 
we understand the possible order of nonlinearity
in accretion disks exhibiting QPOs $k=n-m+4$.
In the next section, we will reproduce several observational results, were not addressed
by other QPO models, what justify the proposition.

\begin{figure}
\epsscale{0.8}
%\hskip-5cm
\plotone{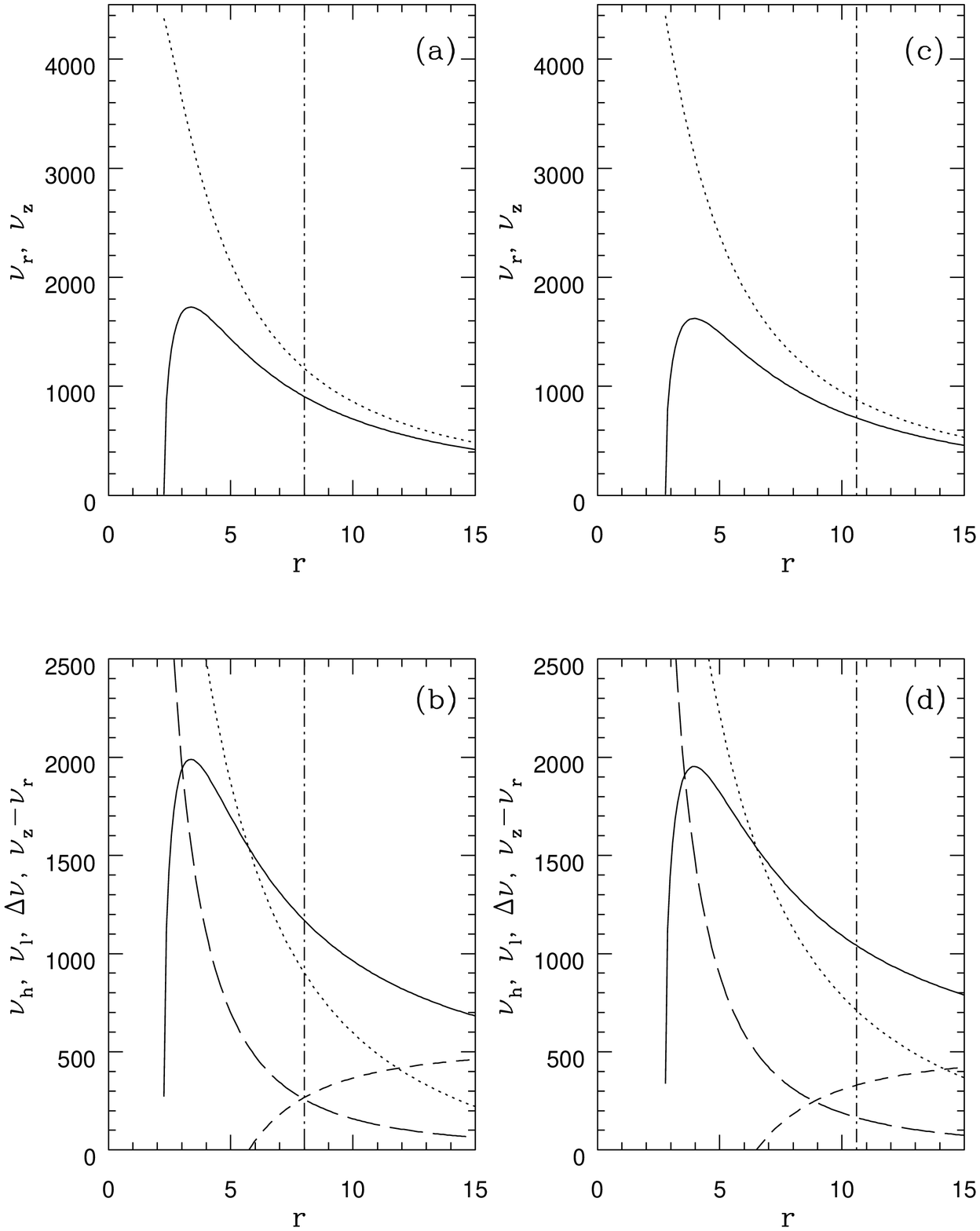}
\caption{ \label{fig1} Variation of
(a) radial (solid curve) and vertical (dotted curve) epicyclic frequencies in Hz as functions of
radial coordinate in unit of $GM/c^2$, (b) higher (solid curve), 
lower (dotted curve) QPO frequencies, their
difference (dashed curve), $\nu_z-\nu_r$ (long-dashed curve) in Hz as functions of 
radial coordinate in unit of $GM/c^2$, 
for the parameter set given in Table 1 for KS~1731-260 when $M=1.1M_\odot$
and $R=16.4$km. 
(c) Same as of (a), (d) same as of (b), but for 4U~1702-429 given in Table 1, 
when $M=M_\odot$ and $R=18.8$km. The vertical dot-dashed curve indicates the
location where the resonance occurs to generate QPO.
}
\end{figure}

In the disk around a neutron star, at an appropriate radius where $\nu_z-\nu_r\sim\nu_s/2$ 
(Kluzniak et al. 2004) and the resonance
is supposed to take place, for $n=m=1$, $\Delta\nu=\nu_h-\nu_l\sim\nu_s/2$, and for 
$n=m=2$, $\Delta\nu\sim\nu_s$. 
$n=1$ corresponds to a nonlinear coupling between the radial
epicyclic mode and the disturbance due to spin of the neutron star, which results in
$\Delta \nu$ locking in the nonlinear regime with $m=1$. On the other hand, $n=2$
corresponds to a linear coupling, which results in $\Delta \nu$ locking in the linear regime with $m=2$.
Hence, for the resonance $n$ must be equal to $m$. 
The resonance with, e.g., $n=2, m=1$ for a neutron star corresponds to a
linear coupling of the epicyclic and the disturbance modes. This results in the locking of $\Delta\nu$
($\sim\nu_s/2$) to be in the nonlinear regime which is forbidden or very weak to observe.
Similarly, $n=3, m=3$ may correspond to higher harmonics of the coupling in the
nonlinear regime which are again expected to be weak to make any observable effects.
It is expected that for a fast rotating neutron star the disturbance occurs 
in the nonlinear regime\footnote{Note that the strength of the disturbance
is expected to be higher for a fast rotating neutron star compared to a 
slow rotating one and, hence, plausibly a nonlinear resonance is in the
former case and a linear resonance in the later one. Indeed observations
showing the apparent relation between QPO and spin frequencies of 
neutron stars support the argument.}. However, both $\nu_r$ and $\nu_z$ and then
$\nu_h$ and $\nu_l$ vary as functions of radial coordinate. Therefore,
$\Delta \nu = \nu_h - \nu_l$ varies as well. We argue that in a small
range of radius, as given in Table 1, the resonance condition remains
satisfied approximately, when $\Delta \nu$ decreases slowly with $\nu_l$, hence
explains observed data. Indeed
observations show $\Delta \nu$ (may vary over $100$Hz for a particular neutron star)
to be of the order of $\nu$ or $\nu/2$ only,
but not to be $\nu$ or $\nu/2$ exactly.
In Fig. \ref{fig1} we describe these facts for stars with $\Delta\nu\sim\nu_s/2$
and $\nu_s$ both. The vertical dot-dashed line indicates the location of the resonance. 

For a black hole, however, in absence of
a magnetosphere, the disturbance and then corresponding coupling between modes may not be
nonlinear and occurs with the condition $\nu_z-\nu_r\lsim\nu_s$, which results 
in the resonance locking at the linear regime
with $n=m=2$ which produces $\Delta \nu\lsim\nu_s$ (sometimes $\sim2\nu_s/3$) as shown in 
Figs. \ref{fig2}a,b. However, if we enforce the resonance to occur at marginally stable circular orbit,
at the very inner edge of an accretion disk where the underlying physics is expected to
be nonlinear than that in the relatively outer edge, then a nonlinear resonance occurs with
$\Delta \nu\lsim\nu_s/2$ (sometime $\sim\nu_s/5$) and $\nu_z-\nu_r\lsim\nu_s$
for $n=m=1$, as shown in Figs. \ref{fig2}c,d.
In the following sections this is discussed in detail. 
%In principle, there are possibilities
%of resonance with other combination of $n$ and $m$ (e.g. $n=2, m=1$) which are expected to be
%weak to observe.

The difference between resonance conditions to form QPOs around black holes and neutron stars might 
be due to difference in their environment. For a neutron star, there is a rotating
magnetosphere imprinting directly the spin frequency on the oscillations of the disk.
However, energy and angular momentum of a black hole with magnetic field
connecting to the surrounding disk can be transferred
to the disk through the variant Blandford-Znajek process. This leads to
the strong resonance, expected to be driven by the disturbance at the spin frequency of 
the black hole, even though the exact physical mechanism remains to be determined.
Another important point to 
note is that for a black hole there is no strict definition of $\nu_s$ corresponding
to its boundary layer related to the spin parameter $a$, while for a neutron star
it is obvious due to presence of its hard surface. Here we assume that the effective boundary 
of a black hole is at marginally stable circular orbit $r_{ms}$ and $\nu_s$ corresponds to 
the frequency of a test particle at $r_{ms}$ [see equations (\ref{bhfreq}), (\ref{bhfreqh})].

From observational data we can determine the spin frequency of a neutron star. 
In computing QPO frequencies from our model, we need to determine the spin parameter $a$.
If we consider a neutron star to be spherical in shape with the equatorial 
radius $R$, spin frequency $\nu_s$, mass $M$, radius of gyration $R_G$, then
moment of inertia and the spin parameter are computed 
\begin{eqnarray}
I=MR_G^2,\,\,\,a=\frac{I\Omega_s}{\frac{GM^2}{c}},
\label{numoi}
\end{eqnarray}
where $\Omega_s=2\pi\nu_s$.
%$G$ is the Newton's gravitation constant and $c$ is speed of light.
We know that for a solid sphere $R_G^2=2\,R^2/5$ and a for hollow sphere $R_G^2=2\,R^2/3$. 
However, the shape of a very fast rotating 
neutron star is expected to be deviated from spherical to ellipsoidal. Moreover, neutron stars are
not expected to be perfect solid bodies. 
%Thus its moment of inertia about vertical axis might be larger with a larger $R_G/R$. 
Hence, in our calculation we choose $0.35\le (R_G/R)^2 \le0.5$ in most cases 
(see e.g. Bejger \& Haensel 2002; Cook, Shapiro \& Teukolsky 1994).

On the other hand, for a black hole of mass $M$, 
$a$ is the most natural quantity what we supply as an input.
%associated with its spin what we consider as a parameter for the calculation. 
Corresponding angular
frequency of a test particle at any radius $r$ in the spacetime around it 
is then given by (e.g. Shapiro \& Teukolsky 1983)
\begin{eqnarray}
\Omega_{BH}=-\frac{g_{\phi t}}{g_{\phi\phi}}=\frac{2a}{r^3+r\,a^2+2a^2}.
\label{bhfreq}
\end{eqnarray}
%which at the horizon $r_+$ reduces to
We now define
\begin{eqnarray}
%\Omega_{BHh}=\Omega_{s}=2\pi\nu_s=\frac{a}{2r_+},\,\,\,r_+=a^2+\sqrt{M^2-a^2}.
\Omega_{s}=2\pi\nu_s=\Omega_{BH}(r=r_{ms})\,\frac{c^3}{GM},
\label{bhfreqh}
\end{eqnarray}
where $r_{ms}$ is marginally stable circular orbit and light inside $r_{ms}$ is practically
not expected to reach us.
Therefore, supplying $a$ we can determine $\nu_s$.

\section{Properties of neutron star QPOs}

For at least seven of neutron stars exhibiting twin kHz QPOs we know the spin frequency
(e.g. Yin et al. 2007; Boutloukos \& Lamb 2008)
In most occasions, QPO frequencies vary with time and 
show that the frequency difference in the pair decreases with the increase of lower
QPO frequency. Moreover, the frequency difference is of the order of half of the 
spin frequency of the star for fast rotators and of the spin frequency itself for slow rotators.
We will show that our model successfully describes all the properties
matching with observed data. In obtaining results, 
we choose primarily $M=1.4M_\odot$, $(R_G/R)^2=0.4$ along with a guess radius $R$; all 
are free parameters. 
However, this choice does not suffice observation mostly. Hence,
subsequently, we discuss results with other values of parameters.
 
\subsection{Fast rotators}

We analyze QPOs from three fast rotating neutron stars whose spin frequencies are known: KS~1731-260
(Smith, Morgan \& Bradt 1997),
4U~1636-53 (Jonker, Mendez, \& van der Klis 2002) 
and 4U~1608-52 (Mendez et al. 1998).
% Lamb \& Miller 2001). 
For various sets of input values of $M$ and $R_G^2/R^2$ we predict radius of the neutron
star, given in Table 1, in reproducing observed data by our theory, depicted in Figs. \ref{fig3}a-c.
KS~1731-260, till date,
has shown to exhibit only one pair of QPO frequency established by
our model easily as depicted in Fig. \ref{fig3}a. However, the resulting $R$ 
is large/unrealistic if mass of the star is considered
to be $M=1.4M_\odot$. If we choose the star not to be a solid sphere or to be an
ellipsoid with $(R_G/R)^2=0.5$ or $0.35$, then the resulting radius decreases to $R=16.4$km which is
in accordance with realistic equations of state (EOS) [Friedman, Ipser \& Parker 1986, hereafter eos1; 
Cook, Shapiro \& Teukolsky 1994, hereafter eos2]. Figures \ref{fig1}a,b
show the resonance to occur at $r_{QPO}=8$ (indicated by the vertical dot-dashed line)
for $n=m=1$ with $\nu_z-\nu_r\sim\Delta\nu\sim\nu_s/2$.

For 4U~1636-53 and 4U~1608-52, on the other hand, twin peaks are observed
in several occasions.
For 4U~1636-53, our theory has an excellent agreement with observation with $\chi^2 \sim 28$,
as shown in Fig. \ref{fig3}b. 
If we exclude only data point falling outside the theoretical
prediction, which does not appear to follow the same trend as other points follow and hence
the significance of its detection appears to be less, then $\chi^2 \sim 6$. 
%Choice of this neutron star not to be a solid sphere or to be an ellipsoid or
%to have mass $M <1.4M_\odot$ reproduces observed data with more realistic radii.
If either $(R_G/R)^2$ or $M$ of the star is considered to be deviated from $0.4$ or $1.4M_\odot$
respectively, then our theory reproduces observed data
in accordance with realistic EOS (eos1, eos2).
While at smaller $\nu_l$-s, results for 4U~1608-52 deviate from observed data, as shown in 
Fig. \ref{fig3}c,
the predicted radii for all the parameter sets, except that with $R=10$km, are well within
the conventional range (eos1, eos2), given in Table 1. 
While the estimated $\chi^2$ for 4U~1608-52 
is large ($> 100$) for the entire set of observed data, it is only
$\sim 8$ when we fit a part of it, discarding remaining data points from the computation
%whose significance of detection appears to be less 
as of 4U~1636-53, by the dot-dashed line shown in Fig. \ref{fig3}c. However, the important point to note is 
that the number of data points in either of the cases is poor which may question 
the reliability of $\chi^2$ values.
Similar resonance as of KS~1731-260 also occurs for these sources as well.

\subsection{Slow rotators}

%Now we analyze, as depicted in Figs. \ref{fig3}d,e, slow rotating neutron stars 
Results for slow rotating neutron stars
whose spin frequencies are known: 4U~1702-429 (Markwardt, Strohmayer \& 
Swank 1999), 4U~1728-34 (van Straaten et al. 2002;
Mendez \& van der Klis 1999), 
are depicted in Figs. \ref{fig3}d,e.
For the sets of input values of $M$ and $(R_G/R)^2$ we predict 
various possible radii of the neutron star, 
given in Table 1. 
If we consider $M=1.4M_\odot$, then the predicted $R$ 
is too large with the choice of a spherical star. If the star is considered to
be an ellipsoid and/or not to be a solid sphere and/or to have mass $M< 1.4M_\odot$, 
then $R$ decreases to conventional values (eos1, eos2).
On the other hand, all the parameter sets for
4U~1702-429, given in Table 1, correspond to high specific angular momentum of the neutron star.
Therefore, we predict that the star 4U~1702-429 can not be a solid sphere and thus corresponding
sets of parameters in rows one and three appear to be ruled out.
Figure \ref{fig3}d shows a perfect agreement of our theory with observation with $\chi^2<1$.
The spin parameter for cases of 4U~1728-34, given in Table 1, ranges $0.5\lsim a\lsim 0.75$. 
The parameter set with $a\sim 0.75$ given in the corresponding row one appears to be ruled out when the star 
is not expected to be a solid sphere and the mass-radius combination does not follow
any realistic EOS. 
While the estimated $\chi^2$ for 4U~1728-34 is large ($> 100$) for the entire set of observed data, 
it is only $\sim 7$ when we fit a part of it, as of 4U~1608-52,
by the dot-dashed line shown in Fig. \ref{fig3}e. However, as mentioned earlier,
the poor number of data points may question the reliability of $\chi^2$ values.
In Figs. \ref{fig1}c,d we describe the resonance and corresponding
formation of QPOs for 4U~1702-429. Clearly, the
resonance occurs in $r_{QPO}=10-11$, indicated by the vertical dot-dashed line, 
for $n=m=2$ when $\nu_z-\nu_r\sim\Delta\nu/2\sim\nu_s/2$.
Similar resonance occurs for other slow rotators.

\subsection{Estimating the spin frequency of Sco~X-1}

The spin frequency of Sco~X-1 is not known yet. 
This source exhibits a pair of kHz QPO frequency with the separation varies in a range $\sim 225-310$ Hz 
(Mendez \& van der Klis 2000).
We show in Fig. \ref{fig3}f the observed variation of the frequency separation as a function of lower QPO
frequency and compare it with that obtained from our model.
Considering all possible sets of input parameters (along with $\nu_s$ which is now unknown)
to obtain the best fit of observations,
we present some of them in Table 1. We find that mass of Sco~X-1 must be less than $1.4M_\odot$
and the sets of inputs with smaller $\nu_s$ and $M$
fit the observed data better and argue that Sco~X-1 is a slow rotator with $\nu_s\sim 280-300$.
For the parameter set with $\nu_s=292$, $\chi^2\sim 66$ when the number of observed 
data points is $39$. However, if we recompute it
discarding data points falling far outside the theoretical prediction, as of 4U~1636-53, 
4U~1608-52, 4U~1728-34, given 
by the dot-long-dashed curve, then $\chi^2\sim 38$ with observed data points $35$,
assuring the excellence of the fitting. 

\subsection{Quality factor}

Quality factor (in short Q-factor) of observed QPO frequencies has been found 
to be as large as $200$. Any viable model for QPO should be able to reproduce
this high Q-factor. Figure \ref{fig4} describes Q-factor of the $775$Hz QPO observed 
from 4U~1636-53, based on our model. The power depends mainly on mechanical resistance 
of the system along with epicyclic frequencies\footnote{Note that the spring constants 
of the underlying oscillators, which vary with radial coordinate, are related to 
the components of epicyclic frequencies.}
and frequency of the neutron star. Following the principles of a forced vibrating 
system of coupled oscillators we obtain the power approximately shown in Fig. \ref{fig4}a.
However, Fig. \ref{fig4}b shows that the power falls off
rapidly away from the resonance radius, because the component of 
radial epicyclic frequency merges with the vertical epicyclic frequency with 
the increase of $r$. Therefore, the resonance condition remains satisfied roughly 
in a (narrow) range of radii, as given in Table 1. The change in amplitude
of the power in that range of radii results in the observed variation of QPO frequency. 
Figure \ref{fig4}b shows that the QPO frequency of 4U~1636-53
varies in the range $r_{QPO}=7.6-8.5$ 
(see Table 1) with amplitude of the power decreases upto one-third of peak amplitude.
Similar features can also be found for other compact sources. 

The value of mechanical resistance $r_m$ of a system controls Q-factor.
From the perturbed Navier-Stokes equation, it is easy to understand that
$r_m$ for an accretion disk, where QPOs occur, is proportional to 
the radial velocity gradient $dv_r/dr$ of the background flow. While for a
Keplerian accretion disk (Shakura \& Sunyaev 1973) 
$dv_r/dr$ is very small, for a sub-Keplerian advective disk (e.g. Mukhopadhyay 2003)
it could be high which is difficult to perturb rendering high mechanical
resistance. In Fig. \ref{fig4}c we show the variation of Q-factor with $r_m$.
It clearly argues that a very large Q-factor is possible for QPOs from
a Keplerian disk.

\section{Properties of black hole QPOs}

The pairs of observed HF QPO from black holes seem to appear at a $3:2$ ratio.
Mass (or possible mass range) of several such black holes, e.g. GRO~J1655-40 
(Orosz \& Bailyn 1997; Shahbaz et al. 1999), XTE~J1550-564 (Orosz et al. 2002), 
GRS~1915+105 (Greiner, Cuby \& McCaughrean 2001),
% Bailyn 2001) 
H1743-322 (Miller et al. 2006) is already estimated from observed data.
However, the spin of them is still not well established. 
We estimate the spin parameter of black holes in describing their observed QPOs by
our model given in Table 2.
%We consider $n=m=1$ and $n=m=2$ both the cases. 

\begin{figure}
\epsscale{0.8}
%\hskip-5cm
\plotone{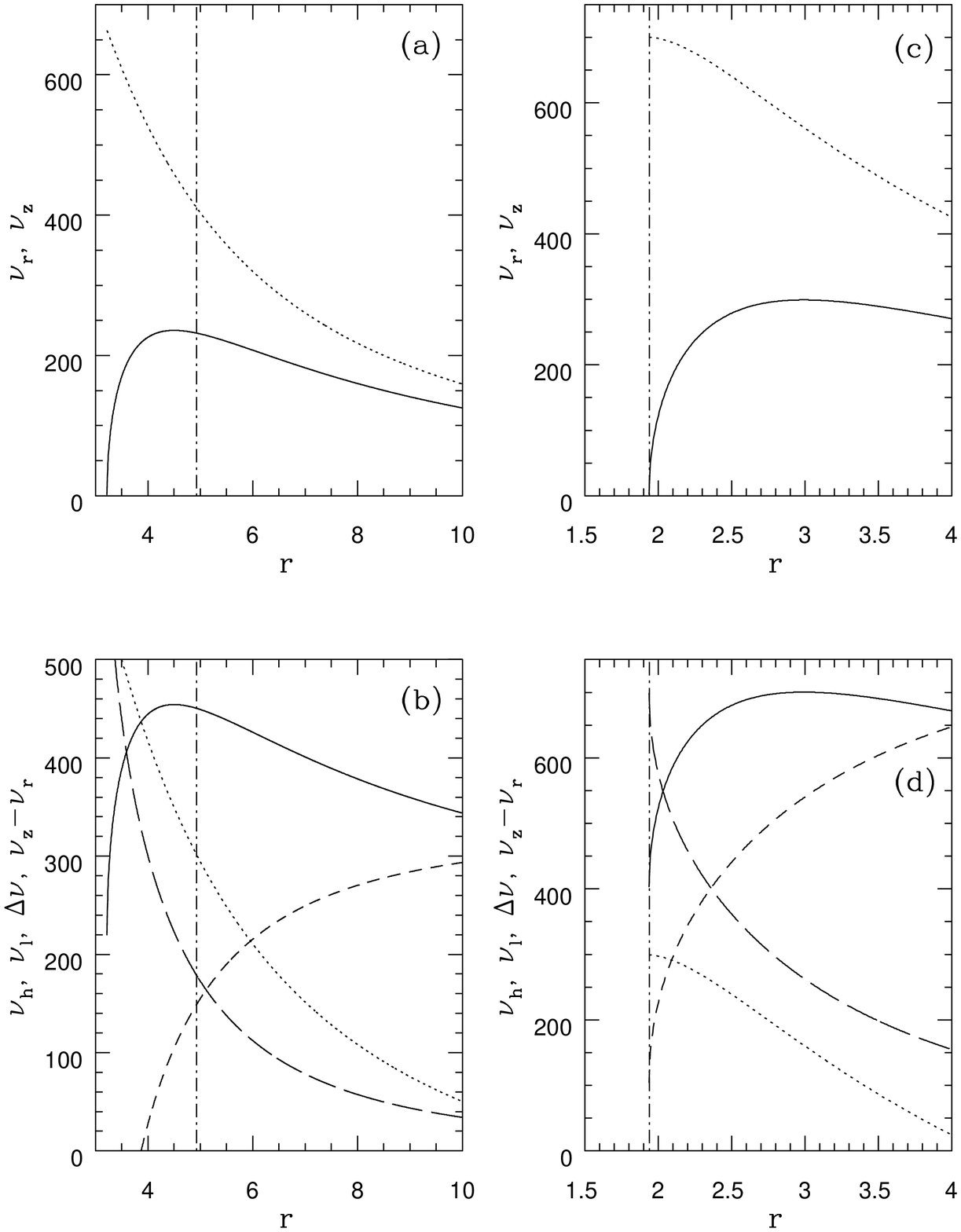}
\caption{ \label{fig2} Variation of
(a) radial (solid curve) and vertical (dotted curve) epicyclic frequencies in Hz as functions of
radial coordinate in unit of $GM/c^2$, (b) higher (solid curve), 
lower (dotted curve) QPO frequencies, their
difference (dashed curve), $\nu_z-\nu_r$ (long-dashed curve) in Hz as functions of 
radial coordinate in unit of $GM/c^2$, for the parameter set given in 
Table 2 for GRO~J1655-40 when $M=6M_\odot$, $n,m=2$.
(c) Same as of (a), (d) same as of (b), but for $M=7.05M_\odot$, $n,m=1$.
The vertical dot-dashed curve indicates the
location where the resonance occurs to generate QPO.
}
\end{figure}

\begin{figure}
\epsscale{0.8}
%\hskip-5cm
\plotone{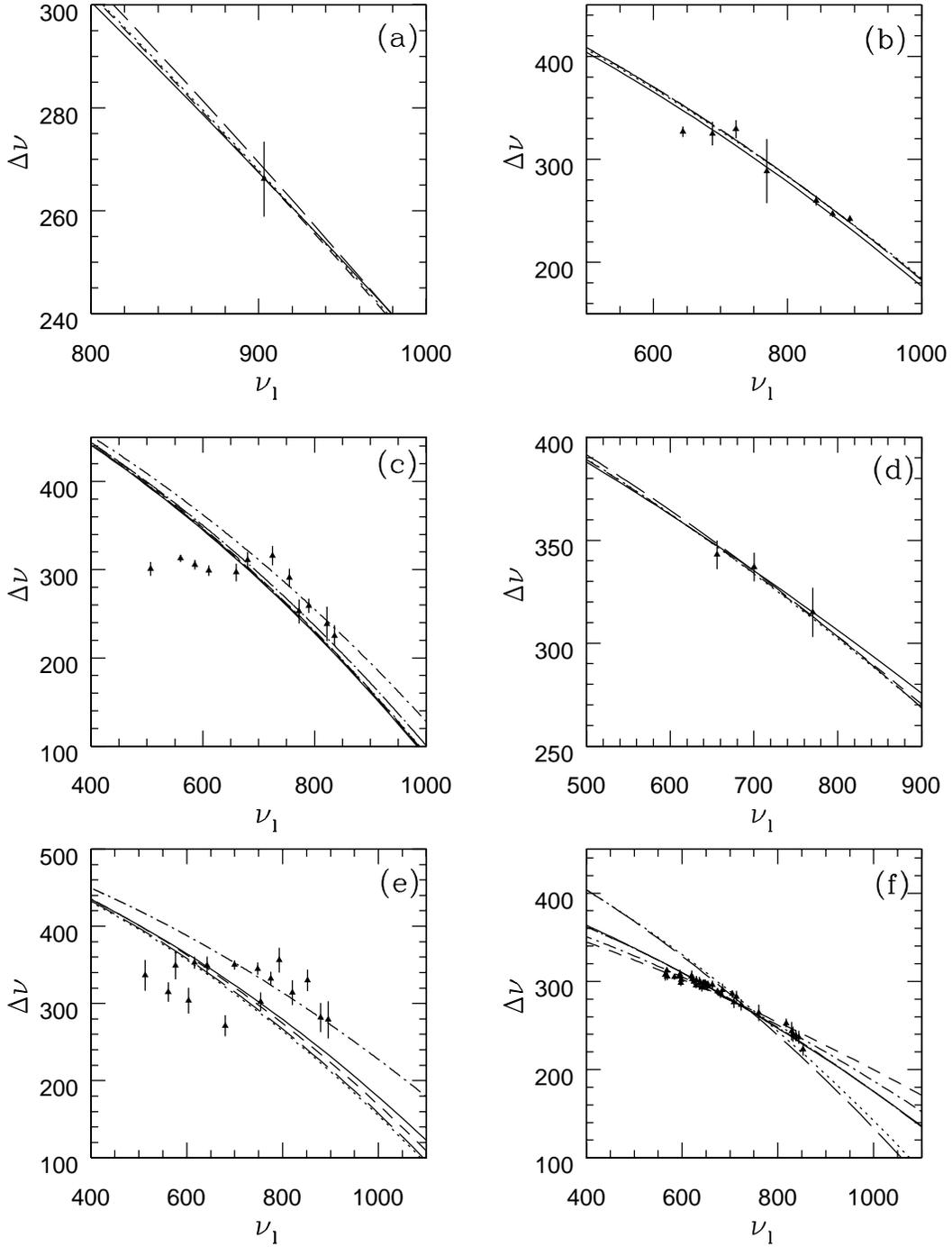}
\caption{ \label{fig3}
Variation of the QPO frequency difference in a pair as a function of lower QPO frequency for
(a) KS~1731-260, (b) 4U~1636-53, (c) 4U~1608-52, (d) 4U~1702-429, (e) 4U~1728-34,
(f) Sco~X-1. Results for parameter sets given in Table 1 from top to bottom row for
a particular neutron star correspond to the solid, dotted, dashed, long-dashed,
dot-dashed (for 4U~1608-52, 4U~1728-34, Sco~X-1 only), dot-long-dashed 
(for 4U~1608-52, Sco~X-1 only) lines. The dot-dashed and dot-long-dashed lines in (c) 
and the dot-dashed line in (e) are
to fit a part of the observed data points discarding the remaining ones.  
The triangles are observed data points along with error bars.
}
\end{figure}

\begin{figure}
\epsscale{0.8}
%\hskip-5cm
\plotone{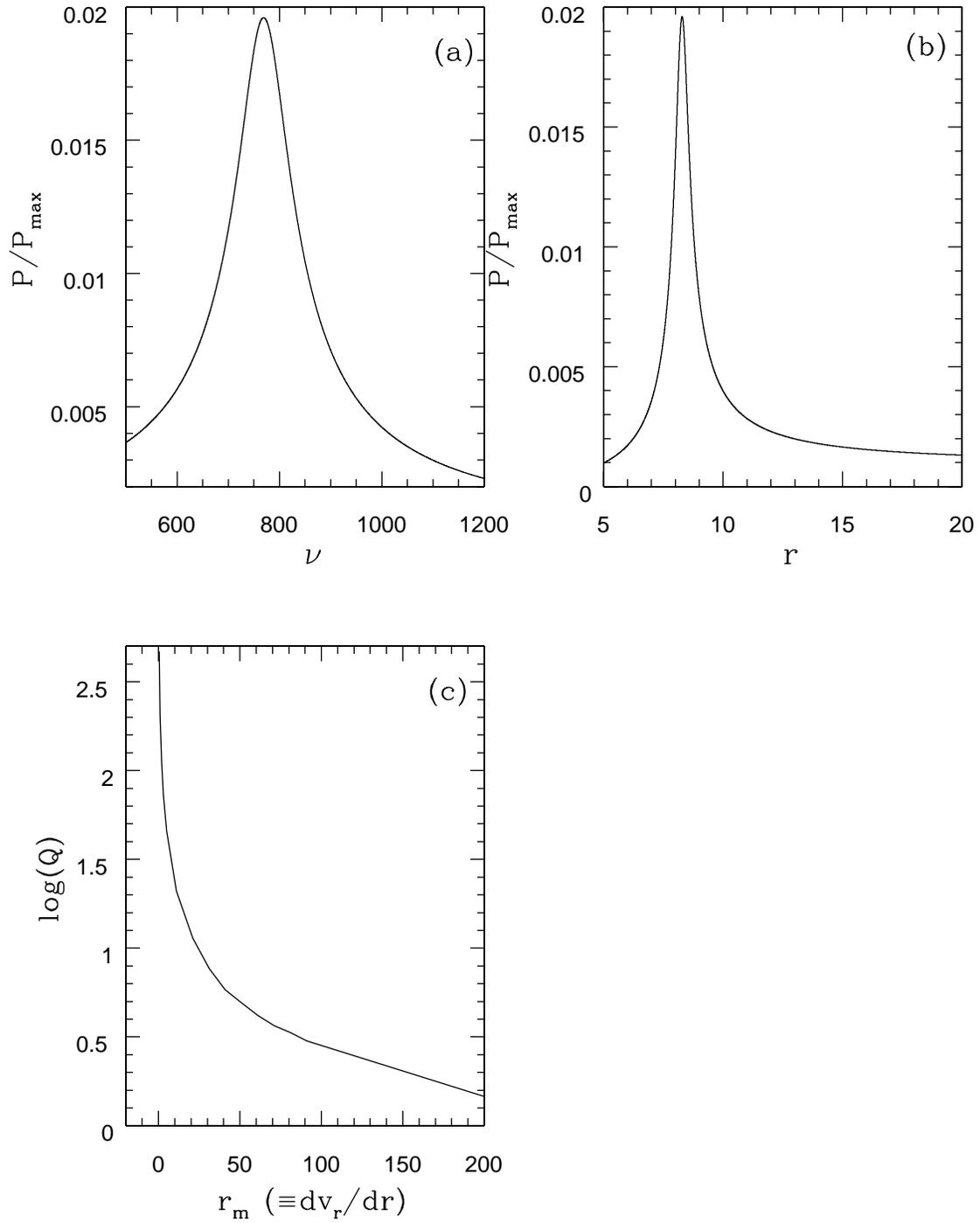}
\caption{ \label{fig4}
(a) Normalized Power-Spectrum for 4U~1636-53 showing $775$Hz QPO, when $M=1.18M_\odot$,
$R=16.8$km, $(R_G/R)^2=0.35$, mechanical resistance $r_m\equiv dv_r/dr\sim 50$sec$^{-1}$.
(b) Variation of the normalized power in (a) as a function of radial coordinate in unit
of $GM/c^2$. (c) Variation of Q-factor as a function of
mechanical resistance of the flow in unit of sec$^{-1}$.
}
\end{figure}

We supply observed $M$ (or range of $M$, whichever is available) and arbitrary values of $a$
as inputs of our model to reproduce observed QPOs. Without knowledge of exact mass for GRO~J1655-40,
XTE~J1550-564 and GRS~1915+105, the estimated range of mass from observed data is supplied for
our computation and we find the corresponding suitable range of input parameter 
$a$ reproducing observed QPOs.
However, mass of H1743-322 appears to be less uncertain with an
estimated value $\sim 11M_\odot$. So we use that value and find a suitable spin parameter.
For clarity, we also choose a (hypothetical) range of mass around $\sim 11M_\odot$ for this
black hole and check in what extent $a$ could vary.
Our theory is able to produce QPOs at a $3:2$ ratio with their observed values
for $n=m=2$ at a radius outside $r_{ms}$. Figures \ref{fig2}a,b describe the
resonance and corresponding formation of QPOs for the black hole GRO~J1655-40
with $M=6M_\odot$ and $\nu_s=218.48$Hz. The vertical dot-dashed line
indicates the location of the resonance. It clearly says that $\nu_z-\nu_r\lsim\nu_s$
at $r_{QPO}=4.93$ and the resonance occurs with $\Delta\nu\lsim\nu_s$ (say, $\sim 2\nu_s/3$,
as shown in Fig. \ref{fig2}b). 
Similar resonance happens for other black holes. The corresponding spin parameter
ranges from $\sim 0.6$ to $\sim 0.8$, depending on mass of the black hole. 

However, if we enforce QPOs 
to happen at $r_{ms}$ strictly by changing mass (within observed range of estimate)
and the spin parameter of a black hole,
then they produce for $n=m=1$ at a higher $a$. Figures \ref{fig2}c,d describe such a
formation of QPOs for the source GRO~J1655-40 with $M=7.05M_\odot$ and $\nu_s=802.48$Hz. 
It again implies that the resonance corresponds to $\nu_z-\nu_r\lsim\nu_s$, but with
$\Delta\nu\lsim\nu_s/2$ (say, $\sim\nu_s/5$, as shown in Fig. \ref{fig2}d). 
However, in this case the required mass of
black holes to reproduce observed QPOs may be out of the range estimated from observed 
data and the corresponding $a$ appears to be similar/same for all four black holes.
Interesting fact to note is that the either case favors the fourth order nonlinearity in accretion disks 
around black holes and none of the parameter sets predict an extremally rotating black hole.
%This, in fact, may be the reason to have their QPO frequencies and masses 
%of same order of magnitude, 
%while we know that $\nu_{h,l}\sim M^{-1}$ (Abramowicz et al. 2004).

In early, employing fully relativistic accretion disk model, 
the spin of black holes was estimated based on a 
spectral analysis of the X-ray continuum (Shafee et al. 2006; McClintock 2006).
While their estimated spin for GRO~J1655-40 is in accordance with ours, for
GRS~1915+105 it is higher ($a>0.98$). However, Middleton et al. (2006), 
based on a simple multicolor disk blackbody including full radiative transfer 
as well as relativistic effects, estimated spin of GRS~1915+105 to be $\sim 0.7$
which tallies with ours. Perhaps, different estimate is due to difference in
method and the disk model used for computation. Indeed, the best model to describe a geometrically
thick accretion disk is still a matter of controversy. In fact, Wang, Ye \& Huang (2007)
showed how does the spin estimate vary with different method.

\section{Summary}

We have described QPOs, particularly the kHz/HF ones, observed from several neutron stars and black holes
(altogether ten compact sources), by a single model and predicted their spin. The model has 
addressed the variation of QPO frequency separation in a pair as a function
of the QPO frequency itself of neutron stars, observed in several occasions. 
We argue that QPO is a result of higher (fourth)
order nonlinear resonance in accretion disks.
The model predicts the most possible set of parameters: $R\sim 15-16$km, $M\sim 1-1.2M_\odot$ for
fast rotators; $R\sim 16-17$km, $M\sim 0.8-1M_\odot$ for slow rotators, when $(R_G/R)^2\sim0.4$, 
that can explain all the observed QPO pairs.
Discussing QPOs for five different neutron stars 
(KS~1731-260, 4U~1636-53, 4U~1608-52, 4U~1702-429, 4U~1728-34) 
whose spin frequencies are known,
we have predicted the spin frequency of Sco~X-1 which is unknown until this work. We argue that
this is a slow rotator.

We have addressed QPOs from four black holes 
(GRO~J1655-40, XTE~J1550-564, H1743-322, GRS~1915+105) 
whose mass (or range of mass) has already been predicted from observed data and QPO pair seems 
to appear at a $3:2$ ratio. 
Based on the present model, we have predicted their spin parameters ($a$-s) which are not 
well established yet.
According to the present model, none of them is an extremally rotating black hole.
As our model explains QPOs observed from several compact objects including their specific properties,
of that in black holes and neutron stars both, 
it favors the idea of QPOs to originate
from a unique mechanism, independent of the nature of compact objects. 
A successful model should be able to reproduce properties globally as the one 
does presented here.

Now the future job should be to perform numerical simulations to establish the resonance 
described here to confirm the model. Importantly, one should compare the energy 
released from the oscillation with that from rest of the disk (and the neutron star
surface, if QPO is from a neutron star) to address visibility 
of oscillations. In this connection, the variation of the disturbance in the
disk due to rotation of the compact object should be employed properly into the 
model to pinpoint the largest possible radius, beyond that QPO does not occur, accurately.

\acknowledgements
This work is partly supported by a project (Grant No. SR/S2/HEP12/2007) funded by Department
of Science and Technology (DST), India. The author would like to thank the referee
for his/her encouraging comment which has helped to prepare the final version of the paper.
Thanks are also directed to Shashikant Gupta of IISc for discussion about $\chi^2$-fitting
and Sudip Bhattacharyya of TIFR for quickly providing references of neutron stars with
known spin and QPO frequencies.

{}

%\clearpage
\begin{table*}[htbp]
\scriptsize
\caption{Physical parameters of neutron stars}
\begin{tabular}{ccccccccccccccccccccccccccccccccccccccc}
\hline
\hline
fast rotator & $\nu_s$ & $M$ & $(R_G/R)^2$ & $n,m$ & $R$ & range of $r_{QPO}$ \\
\hline
\hline
KS~1731-260 & $525.08$ & $1.4$ & $0.4$ & $1$ & $22.01$ & one pair at $6.6$\\
%\hline
%KS~1731-260 & $525.08$ & $1.4$ & $0.6$ & $1$ & $17.97$ & one pair at $6.6$\\
KS~1731-260 & $525.08$ & $1.1$ & $0.5$ & $1$ & $16.4$ & one pair at $8$\\
KS~1731-260 & $525.08$ & $1.1$ & $0.4$ & $1$ & $18.32$ & one pair at $8$\\
KS~1731-260 & $525.08$ & $0.9$ & $0.35$ & $1$ & $16.4$ & one pair at $9.3$\\
\hline
\hline
4U~1636-53 & $581.75$ & $1.4$ & $0.4$ & $1$ & $18.4$ & $6.6-7.4$\\
%\hline
%4U~1636-53 & $581.75$ & $1.4$ & $0.6$ & $1$ & $15.0$ & $6.3-7.6$\\
%4U~1636-53 & $581.75$ & $1.4$ & $0.5$ & $1$ & $16.5$ & $6.7-7.7$\\
%4U~1636-53 & $581.75$ & $1.1$ & $0.4$ & $1$ & $14.3$ & $8.2-9.3$ \\
%4U~1636-53 & $581.75$ & $1.2$ & $0.35$ & $1$ & $16.8$ & $7.7-8.7$ \\
4U~1636-53 & $581.75$ & $1.4$ & $0.49$ & $1$ & $16.8$ & $6.7-7.5$\\
4U~1636-53 & $581.75$ & $1.2$ & $0.4$ & $1$ & $16.0$ & $7.5-8.4$ \\
4U~1636-53 & $581.75$ & $1.18$ & $0.35$ & $1$ & $16.8$ & $7.6-8.5$ \\
\hline
\hline
4U~1608-52 & $619$ & $1.4$ & $0.4$ & $1$ & $15.0$ & $7.0-8.9$\\
%\hline
%4U~1608-52 & $619$ & $1.4$ & $0.6$ & $1$ & $12.5$ & $6.6-7.5$ \\
4U~1608-52 & $619$ & $1.4$ & $0.5$ & $1$ & $13.5$ & $7.0-8.9$ \\
4U~1608-52 & $619$ & $1.1$ & $0.4$ & $1$ & $10.0$ & $8.5-10.7$ \\
4U~1608-52 & $619$ & $1.3$ & $0.35$ & $1$ & $14.5$ & $7.4-9.4$ \\
\hline
4U~1608-52 & $619$ & $1.4$ & $0.45$ & $1$ & $15.1$ & $6.9-8.9$ \\
4U~1608-52 & $619$ & $1.49$ & $0.4$ & $1$ & $16.5$ & $6.6-8.5$ \\
\hline
\hline
slow rotator & $\nu_s$ & $M$ & $(R_G/R)^2$ & $n,m$ & $R$ & range of $r_{QPO}$ \\
\hline
\hline
4U~1702-429 & $330.55$ & $1.4$ & $0.4$ & $2$ & $28.0$ & $7.8-8.6$ \\
%\hline
%4U~1702-429 & $330.55$ & $1.4$ & $0.7$ & $2$ & $21.1$ & $7.8-8.5$ \\
4U~1702-429 & $330.55$ & $1.0$ & $0.5$ & $2$ & $18.8$ & $10.0-11.0$ \\
4U~1702-429 & $330.55$ & $1.1$ & $0.4$ & $2$ & $23.0$ & $9.4-10.3$ \\
4U~1702-429 & $330.55$ & $0.83$ & $0.35$ & $2$ & $18.5$ & $11.6-12.7$ \\
%4U~1702-429 & $330.55$ & $1.1$ & $0.7$ & $2$ & $17.4$ & $9.4-10.2$\\
\hline
\hline
4U~1728-34 & $364.23$ & $1.4$ & $0.4$ & $2$ & $22.5$ & $7.2-9.9$\\
%\hline
%4U~1728-34 & $364.23$ & $1.4$ & $0.6$ & $2$ & $18.0$ & $7.7-9$ \\
4U~1728-34 & $364.23$ & $1.2$ & $0.5$ & $2$ & $16.0$ & $8.1-11.1$ \\
4U~1728-34 & $364.23$ & $1.1$ & $0.4$ & $2$ & $16.5$ & $8.7-11.8$\\
4U~1728-34 & $364.23$ & $1.1$ & $0.35$ & $2$ & $17$ & $8.7-11.8$\\
%4U~1728-34 & $364.23$ & $1.2$ & $0.6$ & $2$ & $15.0$ & $8.7-10.1$\\
\hline
4U~1728-34 & $364.23$ & $1.0$ & $0.4$ & $2$ & $17.5$ & $9.2-12.6$\\
\hline
\hline
 & estimated $\nu_s$ & & & & \\
% & $\nu_s$ & & & & \\
\hline
Sco~X-1 & $300.0$ & $1.4$ & $0.7$ & $2$ & $21.3$ & $7.4-8.8$ \\
Sco~X-1 & $540.0$ & $1.4$ & $0.6$ & $1$ & $15.5$ & $7.0-8.0$ \\
\hline
%Sco~X-1 & $300.0$ & $1.2$ & $0.7$ & $2$ & $18.8$ & $8.3-10.4$\\
Sco~X-1 & $422.0$ & $0.9$ & $0.5$ & $1$ & $17.4$ & $9.9-11.5$\\
Sco~X-1 & $540.0$ & $1.2$ & $0.4$ & $1$ & $16.0$ & $7.9-9.1$\\
\hline
%Sco~X-1 & $280.0$ & $1.1$ & $0.7$ & $2$ & $20.0$ & $8.6-11.4$ \\
Sco~X-1 & $280.0$ & $0.8$ & $0.5$ & $2$ & $17.5$ & $11.4-13.3$ \\
Sco~X-1 & $292.0$ & $0.81$ & $0.35$ & $2$ & $18.6$ & $11.3-13.2$ \\
\hline
\hline
\end{tabular}
\tablecomments{$\nu_s$ is given in unit of Hz, $M$ in $M_\odot$, $R$ in km, 
radial coordinate where QPO occurs, $r_{QPO}$, in unit of $GM/c^2$.}
\end{table*}

%\clearpage
\begin{table*}[htbp]
\scriptsize
\caption{Physical parameters of black holes}
\begin{tabular}{ccccccccccccccccccccccccccccccccccccccc}
\hline
\hline
black hole & $M$ & estimated $a$ & $\nu_h$ & $\nu_l$
%\multicolumn{2}{c|}{observed} \\
%\cline{2-10}
& $r_{QPO}$ & $\Delta r$ & $n,m$ \\
& & & theory/observation & theory/observation & & \\
\hline
\hline
%GRO~J1655-40 & $6.3$ & $0.75$ & $449.74/450$ & $297.96/300$ & $4.74$ & $1.58$ & $2$\\
GRO~J1655-40 & $6-7$ & $0.737-0.778$ & $450/450$ & $300/300$ & $4.93-4.25$ & $1.71-1.23$ & $2$\\
%GRO~J1655-40 & $4-9$ & $0.623-0.853$ & $450$ & $300$ & $7-3.04$ & $3.27-0.43$ & $2$\\
GRO~J1655-40 & $7.05$ & $0.95$ & $451.31/450$ & $299.04/300$ & $1.94$ & $0$ & $1$\\
\hline
%XTE~J1550-564 & $10.2$ & $0.75$ & $277.78/276$ & $184.03/184$ & $4.74$ & $1.58$ & $2$ \\
XTE~J1550-564 & $8-11$ & $0.682-0.768$ & $276/276$ & $184/184$ & $5.91-4.4$ & $2.44-1.33$ & $2$ \\
%XTE~J1550-564 & $6-14$ & $0.603-0.837$ & $276$ & $184$ & $7.5-3.3$ & $3.68-0.6$ & $2$ \\
XTE~J1550-564 & $11.5$ & $0.95$ & $276.67/276$ & $183.32/184$ & $1.94$ & $0$ & $1$ \\
\hline
H1743-322 & $11.3$ & $0.74$ & $242.39/242$ & $165.89/166$ & $4.81$ & $1.6$ & $2$ \\
H1743-322 & $8-15$ & $0.644-0.82$ & $242/242$ & $166/166$ & $6.5-3.5$ & $2.86-0.7$ & $2$ \\
H1743-322 & $12.7$ & $0.95$ & $242.46/242$ & $166.05/166$ & $1.94$ & $0$ & $1$ \\
\hline
GRS~1915+105 & $10-20$ & $0.606-0.797$ & $168/168$ & $113/113$ & $7.38-3.9$ & $3.58-0.98$ & $2$\\
%GRS~1915+105 & $16.4$ & $0.74$ & $167.01/168$ & $114.3/113$ & $4.81$ & $1.6$ & $2$\\
GRS~1915+105 & $18.4$ & $0.95$ & $167.35/168$ & $114.61/113$ & $1.94$ & $0$ & $1$\\
\hline
\hline
\end{tabular}
\tablecomments
{$\nu_{l,h}$ are given in unit of Hz, $M$ in $M_\odot$,
$r_{QPO}$ and its distance from marginally stable orbit, $\Delta r$,
are expressed in unit of $GM/c^2$.}
\end{table*}

\clearpage

\end{document}